\documentclass[11pt,twoside]{article}
\usepackage{asp2010}

 \newcommand{\mics}{$\mu$m~}

\def\tex {\ifmmode{{T}_{\rm ex}}\else{$T_{\rm ex}$}\fi}
\def\tmb {\ifmmode{{T}_{\rm mb}}\else{$T_{\rm mb}$}\fi}
\def\ci     {\ifmmode{{\rm C}{\rm \small I}}\else{C\ts {\scriptsize I}}\fi}
\def\hi     {\ifmmode{{\rm H}{\rm \small I}}\else{H\ts {\scriptsize I}}\fi}
\def\hh     {\ifmmode{{\rm H}_2}\else{H$_2$}\fi}

\def\ts     {\thinspace}
\def\kms    {\ifmmode{{\rm \ts km\ts s}^{-1}}\else{\ts km\ts s$^{-1}$}\fi}
\def\msol   {\ifmmode{{\rm M}_{\odot}}\else{M$_{\odot}$}\fi}
\def\lsol   {\ifmmode{{\rm L}_{\odot}}\else{L$_{\odot}$}\fi}
\def\zsol   {\ifmmode{{\rm Z}_{\odot}}\else{Z$_{\odot}$}\fi}
\def\etal   {{\rm et\ts al.\ts}}

\resetcounters

\begin{document}

\title{Molecular gas in high redshift galaxies}
\author{Francoise Combes,$^1$}
\affil{$^1$Observatoire de Paris, LERMA, CNRS, 61 Av de l'Observatoire, 75014, Paris, France}

\begin{abstract}
Recent observations with the IRAM instruments have allowed to explore the 
star formation efficiency in galaxies as a function of redshift, in detecting and mapping
their molecular gas. Some galaxies stand on what is called the ``main sequence'', forming
stars with a rate that can be sustained over time-scales of 1 Gyr, some are starbursts, with 
a much shorter depletion time.  Star formation was more active in the past,
partly because galaxies contained a larger gas fraction, and also because the star formation
efficiency was higher. The global Kennicutt-Schmidt relation was however similar until 
z$\sim$ 2.5. Magnification by gravitational lenses have been used to explore in details
galaxies at higher redshift up to 6. Herschel has discovered many of these candidates, and 
their redshift has been determined through the CO lines. ALMA is beginning to extend
considerably these redshift searches, with  its broad-band receivers, for a large range
of objects too obscured to be seen in the optical. 
\end{abstract}

\section{Main sequence and starburst galaxies}

Star formation occurs in galaxies with different speeds of interstellar gas consumption.
Two main modes can be distinguished, either a quiescent mode or "main sequence" mode,
where the gas depletion time-scale is larger than 1 Gyr,
and a starburst mode, for which the depletion time-scale is 100 Myr or smaller.
 Locally, the average gas depletion time-scale for the main sequence star-forming
galaxies is 2.35 Gyr (Bigiel \etal 2011). Starbursts are thought to be a transient phase, triggered
by an external dynamical event, such as a galaxy interaction, a minor or major merger,
or gas accretion followed by violent gas flows towards the center.

\begin{figure}[!h]
\centering
\resizebox{\hsize}{!}{\includegraphics{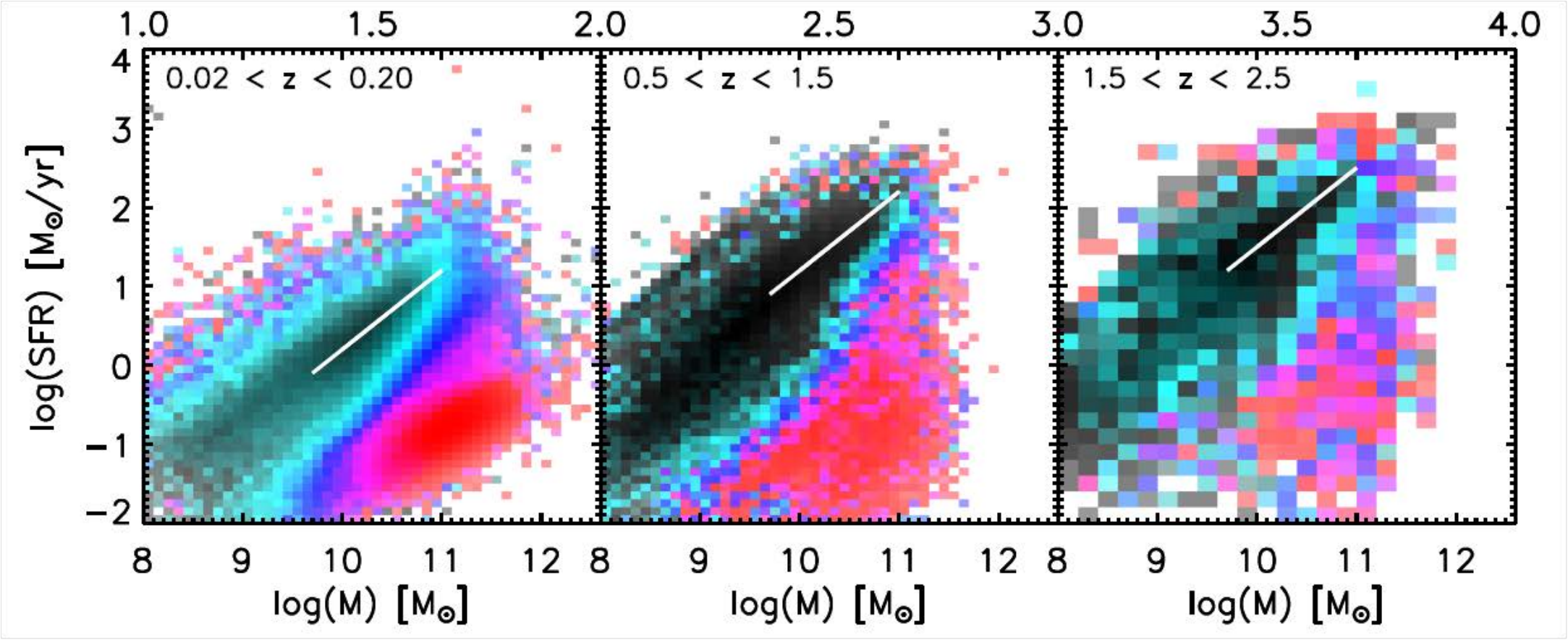}}
\caption{ Statistics on about 800 000 galaxies of the SFR versus stellar mass,
for three ranges of redshifts (from Wuyts \etal 2011). There is a 
correlation between SFR, structure (Sersic index) and stellar populations, since
z $\sim$ 2.5 (or lookback time of 11 Gyr). About 
90\% of cosmic star formation occurs on the  main sequence (white line of slope 1).
Starbursts are located at the top envelope of the main sequence.}
\label{fig1}
\end{figure}

The main sequence of star formation has been identified at least up to 
redshift 2.5 (Noeske \etal 2007; Elbaz \etal 2007; Daddi \etal 2007),
and the rate of star formation on this sequence increases steadily with redshift,
like $(1+z)^m$, with $m\sim 3$. On the main sequence, the star formation rate (SFR) is
roughly proportional to the stellar mass M$_*$.
Wuyts \etal (2011) have studied the statistics of hundreds of thousands of local galaxies from the SDSS,
and high-z galaxies from HST imaging together with multiwavelength photometry and spectroscopy,
and concluded that, at a given stellar mass, there is a large scatter in SFR, from the red 
sequence of quenched galaxies to the main sequence corresponding to the blue ones (cf Figure
\ref{fig1}). The color from stellar populations is correlated
to the morphology, or luminosity profile (stellar density structure) obtained through
the Sersic index derivation.
Star forming galaxies on the main sequence are exponential disks,
while red quiescent systems have a de Vaucouleurs profile.

\section{ULIRGs at intermediate redshift}

To explore the main parameters causing the drop by an order of magnitude of the 
cosmic SFR between z=1 and z=0, we first study the most extreme objects, the 
ultraluminous (ULIRG), which are usually starbursting galaxies, triggered
by interactions and mergers.  About 70 objects were selected in the northern hemisphere
through their FIR luminosity, and observed in the CO lines with IRAM-30m
(Combes \etal 2011, 2013). Their redshift
range is 0.2-1, to fill the gap already noticed by Greve \etal (2005).

\begin{figure}[!h]
\centering
\resizebox{\hsize}{!}{\includegraphics{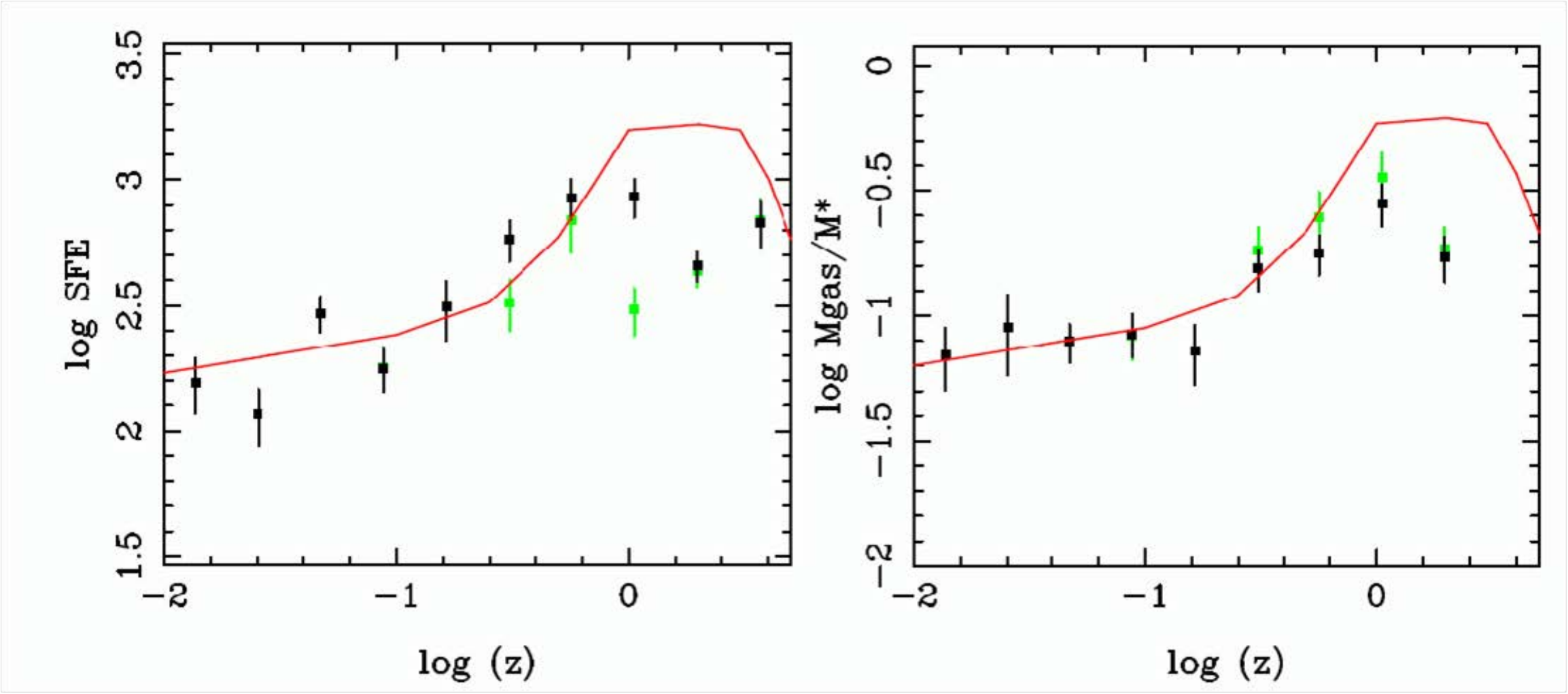}}
\caption{Evolution with redshift of averaged quantities, SFE to the left, and gas-to-stellar mass ratio
to the right, for ULIRGs (Combes \etal 2013). 
The average of only detected points is plotted in green, and with the 3$\sigma$ upper limits in black
(for high-z samples only). The error bars are the statistical ones following the square root of the number of points averaged.
The red line is adapted from the cosmic star formation density from Hopkins \& Beacom (2006).
}
\label{fig2}
\end{figure}

We detected 60\% of the objects with 0.2 $<$ z $<$ 0.6, and 38\% in the 
redshift range 0.6-1.0. According to the excitation of the gas, determined in a 
few of the objects with several high-J CO lines, and also to the dust mass derived
from the far-infrared, we adopt a low CO-to-H$_2$ conversion factor, as proposed
by Solomon \etal (1997) for ULIRGs.
We define the star formation efficiency (SFE) by the ratio of far infrared luminosity, a 
proxy for the star formation rate,
to the CO line luminosity, a proxy for the molecular gas content.
The SFE is found to increase with redshift, by about a factor 3, when 
all detected ULIRG are taken into account. The stellar mass of all the objects
has been derived from modeling their SED in the optical and near infrared.
It is then possible to study the evolution of the gas mass fraction with redshift.
Figure \ref{fig2} shows that both variations of SFE and gas-to-stellar mass ratios
are comparable, they both evolve in the sense of the cosmic SFR density.
Their amplitude is howeveer shallower, they both
increase by a factor $\sim$ 3 between z=0 and 1.
It can be concluded that both factors are required to explain the total
SFR variation.

\section{Main sequence galaxies at z=1-2}

We have undertaken a survey in the CO(3-2) line 
with the IRAM interferometer of about 50 massive star forming galaxies,
selected from their stellar mass and their star formation rate, in the AEGIS survey (Tacconi \etal 2010, 2013). 
Although these galaxies are forming stars at a high rate, they are still in the blue main
sequence at their respective redshift of 1.2 and 2.3, given the average SFR increase with
redshift.

\begin{figure}[!h]
\centering
\resizebox{\hsize}{!}{\includegraphics{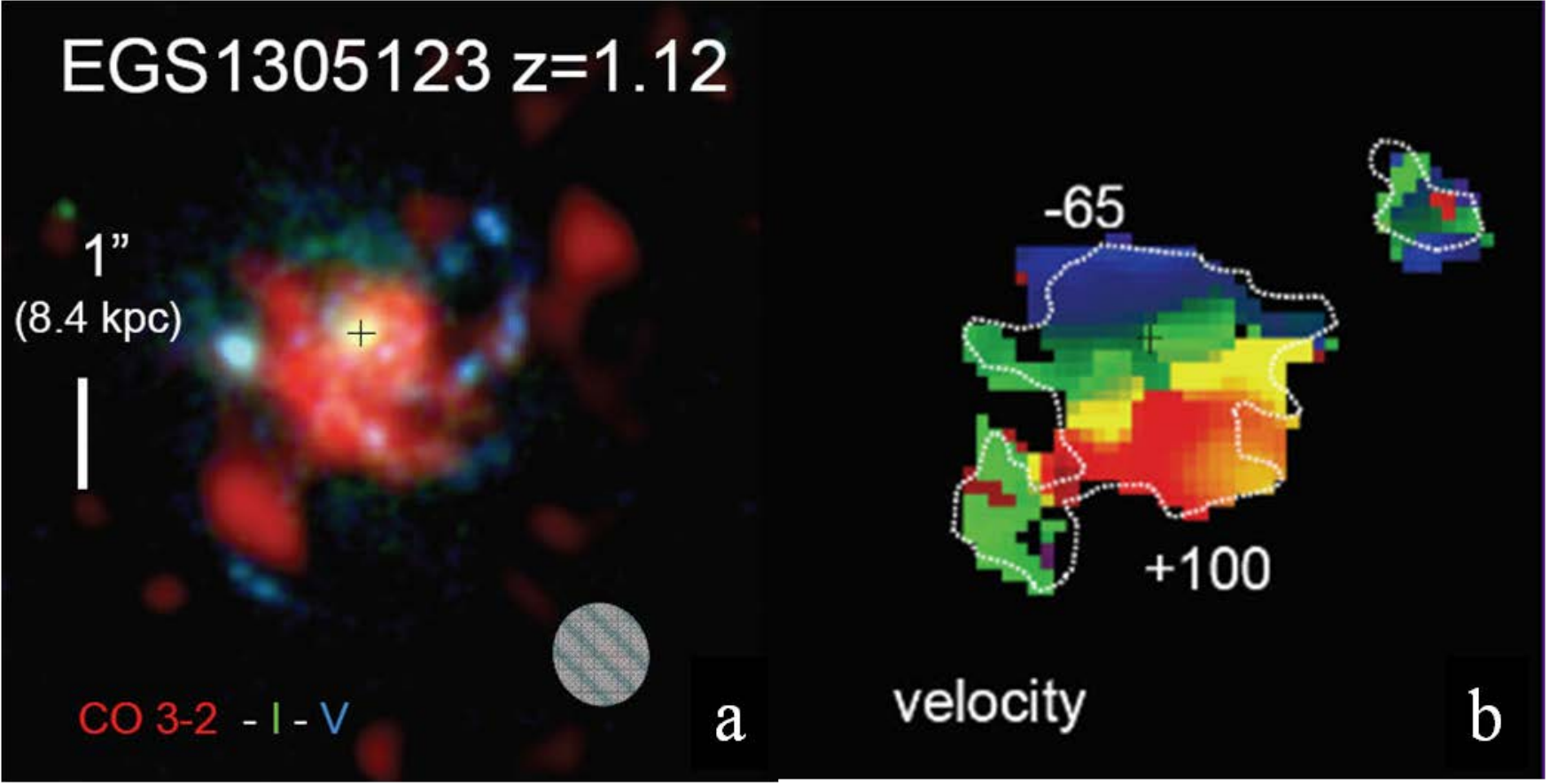}}
\caption{One of the object of the z=1.2 sample: the galaxy EGS1305123 at redshift z=1.12.
{\it Left:} Superposition of the map of the CO line (red, obtained with the IRAM interferometer), with the 
I-band (green) and V-band (blue) images obtained with the Hubble telescope. 
The CO(3-2) line is redshifted at 2mm wavelength, and is mapped at a spatial resolution of 
0.6"x0.7" (beam indicated by the grey hatched ellipse).
{\it Right:} Velocity field of the galaxy in the CO line. From Tacconi \etal (2010)
}
\label{fig3}
\end{figure}

Almost all these high-mass galaxies are detected in CO, and the main result is their increased
gas fraction. While local galaxies have about 5\% of their baryons in gas
(e.g. Saintonge \etal 2011), the average gas content is $\sim$34\% and 44\% 
at z=1.2 and 2.3 respectively (Tacconi \etal 2010). These galaxies have relatively
regular morphologies, those of large spirals not experiencing mergers (or very minor ones).
Their gas velocity fields show regular rotation, although with higher velocity dispersion
than at z=0 (cf Figure \ref{fig3}). Their gas depletion time-scale is slightly below 1 Gyr, meaning that their
star forming duty cycle is relatively short, or they are constantly replenished in gas,
through external accretion from cosmic filaments (e.g. Keres \etal 2005, Dekel \& Birnboim 2006).

Some of the objects are strong enough in the CO line that it was possible to resolve
their disk with higher spatial resolution, of the order of one arcsec or below.
Although it is not possible to identify the bright clumps seen in the HST images at 0.1'' 
resolution, the spectroscopy both in [OII] and in CO allow to separate clumps
with their different kinematics along a given line of sight, and to attempt the derivation
of the resolved Kennicutt-Schmidt relation. The correlation has a lot of scatter,
but reveals an average depletion time-scale for the clumps of $\sim$ 1.5 Gyr (Freundlich \etal 2013).

\section{Lensed galaxies up to z=6}

To be able to detect and map galaxies at even higher redshift, most studies have used the
magnification of gravitational lenses. The HLS key project on Herschel (PI E. Egami) has 
targeted relatively nearby galaxy clusters, to find background high-z candidate galaxies.
The selection is based on objects whose peak of dust emission is redshifted from
100\mics to 500\mics in the SPIRE band.  Several objects were searched
for CO lines, to determine their redshifts. One of the strongest galaxies has
been found behind the cluster Abell 773, and many molecular lines have been detected
in the redshift search, including H$_2$O, CI, [CII], [NII]205\mics (Combes \etal 2012).

Follow up observations with SMA and PdBI have resolved the strong lensing images,
the magnification is about 11, and the source presents two arcs, forming an incomplete ring.
 The velocity profile reveals at least two different components (two interacting galaxies?)
 and the [CII] line map indicates that the blue and red components
are distributed in complementary arcs, forming an interrupted Einstein ring (Boone \etal 2013, in prep).
The high signal-to-noise spectra in CI and CO(7-6) at about the same frequency
have allowed to put
constraints on variation of fundamental constants, in particular
the proton-to-electron mass ratio (relative variations below 2 10$^{-5}$, Levshakov \etal 2012).

Rawle \etal (2013, in prep) show that the [NII] to [CII] ratio is 0.043, similar to that in nearby objects.
The ratio can be used as a metallicity diagnostic, as shown by Nagao \etal (2012)
with ALMA. This diagnostic is precious, for dust-enshrouded objects at z$>$3,
since there are no available diagnostics in optical lines from the ground.

\section{ALMA mapping at z=0}

\begin{figure}[!h]
\centering
\resizebox{\hsize}{!}{\includegraphics{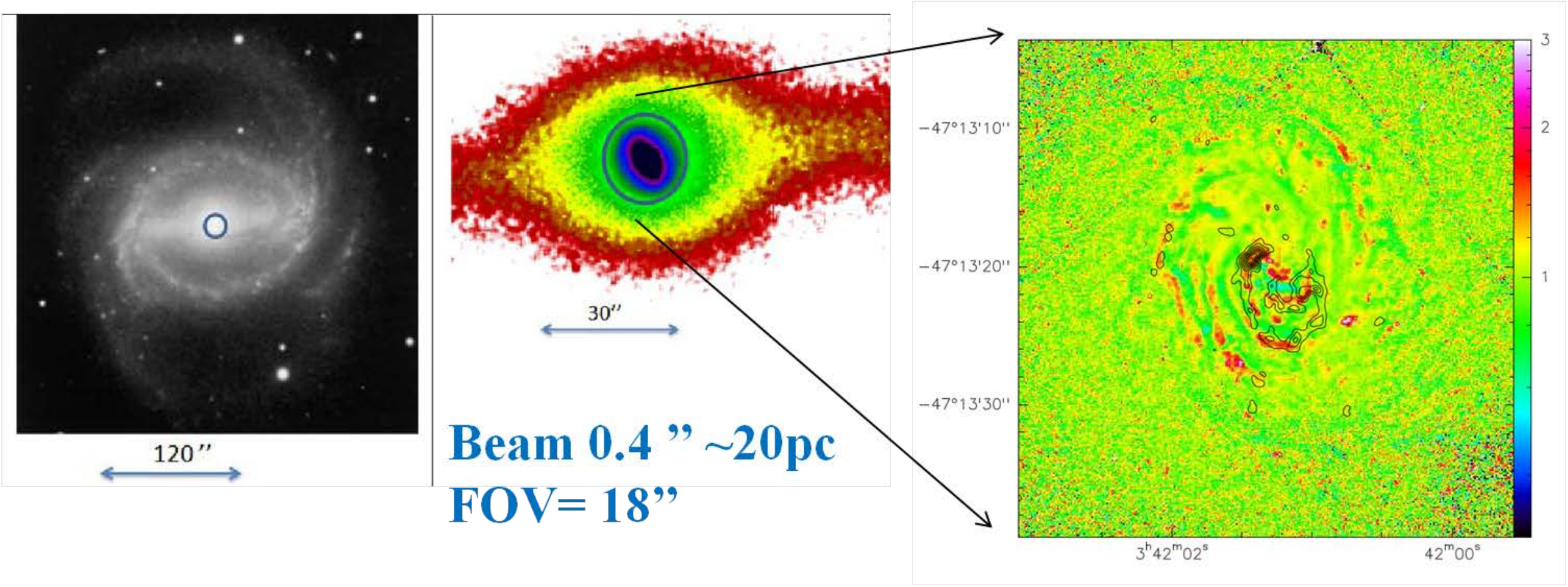}}
\caption{ALMA CO(3-2) map of the nearby barred Seyfert 2 galaxy NGC 1433.
Left is a large-scale visible image of the galaxy, Middle is a zoom in the near infrared, to
better show the nuclear bar, inside the large bar, Right is the HST blue image (colours)
with the CO(3-2) contours superposed, from ALMA (Combes \etal 2013, in prep).
}
\label{fig4}
\end{figure}

Coming back to local galaxies, we have been able to use ALMA in cycle 0 to explore
two Seyfert galaxies, and their molecular gas fueling mechanisms.
NGC~1433 is a Seyfert 2 called the Lord of rings (Buta \& Combes 1996), since it shows 
clearly at least 3 rings corresponding to resonances: outer ring at OLR,
inner ring at UHR near corotation, and a nuclear ring at the ILR of the bar.
 The field of view of the ALMA observations in band 7 is 18'' only, and corresponds
to the nuclear bar, inside the nuclear ring. Figure \ref{fig4} reveals that the CO(3-2)
emission follows quite well the spiral structure seen in dust lanes with HST, and 
corresponding to an inner spiral inside the nuclear bar. The spatial resolution
is 0.4  arcsecond or $\sim$ 20pc. Neither strong HCO$^+$ nor HCN is seen in this galaxy. 
On the contrary, these molecules have strong emission in the Seyfert 1 NGC 1566. 
The HCO$^+$/CO ratio is surprisingly strong, as it is in M82 (Seaquist \& Frayer 2000).
This might indicate a predominance of PDR, or cosmic ray heating.
Contrary to expectations, the Seyfert 1 has a much denser and richer molecular
disk, at 20pc resolution, possibly corresponding to a molcular torus.

\section{Perspectives}

Both continuum dust emission and CO lines are increasingly observed at high redshifts,
and ALMA coming on line will make considerable breakthroughs in this domain.
It will be possible to follow the main sequence of star forming galaxies, not only
at very high masses, nor high SFR. More normal galaxies, with 
modest SFR, and lower stellar masses could be reached, revealing an unbiased main sequence.
The cosmic evolution of the gas content of galaxies
will be unveiled, and the efficiency of star formation established across the
Hubble time. With high spatial
resolution, it will be possible to check whether the Kennicutt-Schmidt relation is still valid
at high-z, an information crucial for models of galaxy formation and evolution. 
 The mechanisms responsible for triggering SFR, or for quenching it, will be studied in detail
with resolved maps, and taking into account environmental effects.

\acknowledgements Many thanks to the organizers for the invitation to such a beautiful and exciting conference.


\end{document}